# GhostAPI-V: Evaluating Audited Diagnostics for Internal API Migration Repair

Jingtao Zhang
Georgia Institute of Technology
United States
jzhang957@gatech.edu

Zeming Liu
Brown University
United States

Haorui Gao
New York University
United States

***Abstract*—Internal API migrations require callers to adopt target-version usages while removing incompatible legacy usages. GhostAPI-V represents these obligations as human-audited temporal contracts and supplies candidate-specific repair diagnostics, with correctness assessed by independent behavioral tests. We evaluate 48 internal migrations from eight Python repositories using Claude Code and Codex, with two initial runs per agent and task. Five revision arms branch from the same 192 initial candidates, controlling generation budget, version evidence, and development-test feedback. Hidden-test success is 120/192 (62.5%) for raw version-difference feedback, 130/192 (67.7%) for required-only contracts, and 148/192 (77.1%) for full bidirectional contracts. Full feedback improves over raw version-difference feedback and required-only feedback by 14.6 and 9.4 percentage points, respectively. The package-level effect is positive in all eight repositories and survives the reported repository sensitivity analyses; the forbidden increment is analysis-sensitive. Post-freeze challenges expose symbolic–behavioral disagreement. The results support the complete audited package beyond the same raw evidence and test failures, while leaving human interpretation, obligation wording, and checking entangled.**

## I. Introduction

An internal API migration can require several coordinated edits. Renaming an import may be insufficient if a caller must also instantiate a new provider, change an argument convention, and remove an obsolete registration hook. Repository-level coding agents can retrieve individually plausible usages from different versions and still leave this coordinated transition incomplete. The validity boundary is the target repository commit, not whether an identifier has existed somewhere in the project's history.

Flask's move from JSON encoder hooks to a provider object illustrates the distinction. A candidate can instantiate the target provider while retaining an obsolete encoder assignment; another can remove the obsolete hook without installing the provider. The first violates a source-side prohibition, the second a target-side obligation. A bidirectional diagnostic describes both defects, while a target-only checklist expresses only the second. Whether this extra diagnostic actually improves executable patches is an empirical question rather than a consequence of the representation.

Repository retrieval is valuable [1], [2], and evolving-API benchmarks identify version conflicts as a substantial generation problem [3], [4]. Cross-version graphs and migration fact checking are already studied [5], [6]. Our contribution is narrower: audited, commit-conditioned obligations for a repository's own callers, checked after generation and supplied as one repair diagnostic. We do not claim a new general graph representation, automatic oracle discovery, or a proof of patch correctness.

A key alternative explanation was that any second attempt, test log, or extra source diff could produce the same gain. The evaluation therefore compares all revision methods from identical initial candidates, with an independent behavioral endpoint. A direct-evidence control receives the same provider evolution evidence used to construct the contracts. A required-only control removes just the forbidden component. These comparisons test the value of the diagnostic package and its forbidden component without confounding them with additional rollouts or privileged access to a reference patch.

We contribute a controlled evaluation against direct version evidence and required-only feedback, plus an analysis of symbolic–behavioral disagreement. Events and templates from the formative pilot were excluded.

We address two research questions. **RQ1:** Does full temporal feedback improve independently measured migration success over the same raw version evidence and development-test feedback? **RQ2:** Does the forbidden component provide an incremental benefit, and how often do symbolic diagnostics disagree with behavior? We answer both questions through controlled contrasts; a before/after gain alone does not identify the diagnostic contribution.

## II. Audited Temporal Contracts

### A. Representation and Scope

Let $s$ and $t$ denote source and target revisions of repository $r$. The editable caller is initialized to its pre-migration state, while the provider implementation and dependencies correspond to $t$. A task-specific temporal contract is

$$E_{s\to t} = (R_{s\to t}, F_{s\to t}, V_t), \tag{1}$$

where $R$ describes required target usages, $F$ describes forbidden source usages, and $V_t$=($V_{dev}$,$V_{hidden}$) contains the target-version development and hidden behavioral checks. Contract authors see neither $V_{hidden}$ nor its outcomes. "Temporal" means conditioned on this source/target transition; "bidirectional" means required and forbidden usages, not reverse migration. We abbreviate $R$ and $F$ for a fixed transition. For materialized candidate code $p$, a usage matcher $m$ defines symbolic satisfaction:

$$\mathrm{sat}(p, E) = \bigwedge_{u\in R} m(u, p) \ \wedge \bigwedge_{v\in F} \neg m(v, p). \tag{2}$$

The experiment uses the prototype's normalized usage matcher. This lexical check does not perform semantic analysis. Whitespace is normalized and definition/decorator patterns compare normalized lines. Here $p$ is the complete content of the predeclared task-scoped caller files after applying the submitted patch, including unchanged code. Matching operates on the resulting files rather than deleted lines or response prose. Correct alternative spellings can still fail $m$, and irrelevant mentions can pass it.

The repair diagnostic consists of

$$\begin{aligned} M(p) &= \{u \in R : m(u,p) = 0\}, \\ L(p) &= \{v \in F : m(v,p) = 1\}. \end{aligned} \tag{3}$$

Each item includes the relevant source/target provider anchor and a short explanation. For the provider example, $M(p)$ can name a missing provider installation and $L(p)$ a remaining encoder-hook assignment. The agent receives the original request, its previous candidate, and the allowed feedback and returns one complete revised patch. The matcher supplies advice, not the acceptance criterion.

### *B. Evidence-Limited Construction*

A curator first identifies a historical provider change and associated caller migration. The target caller patch is sealed as a reference for task admission, not supplied to contract authors or agents. The contract author sees only the maintenance request, the reconstructed legacy caller, and the provider-only source/target evidence packet. Required and forbidden usages are proposed from that packet and audited for traceability, applicability, and excessive restriction to one implementation. This study evaluates audited contracts; precision of a fully automatic synthesizer is outside its claims.

For every admitted task, the contract author justified each expectation from the packet. For example, a replacement provider interface can justify an installation requirement, while removal of the legacy hook justifies its prohibition. No obligation was copied from the sealed target caller patch or hidden tests. The direct-diff control received the identical packet, including the source and target provider excerpts from which these explanations were derived. The advantage tested is the complete package of human interpretation, obligation wording, and candidate-specific diagnostics. It does not isolate checking from supplying the same audited checklist without checking. Human curation was an explicit system cost. The contract author and auditor were distinct; both were blind to the sealed reference and hidden tests.

### *C. Independent Execution Gate*

Let $B_{s\to t}$ be the target-provider checkout with legacy caller hunks restored. Development checks $V_{dev}$ informed repair in designated arms, while hidden checks $V_{hidden}$ were disjoint assertions and fixtures frozen before model runs. Let $p_\Delta$ be the caller materialized by applying the net patch $\Delta$ to $B_{s\to t}$. The primary outcome is

$$Y(\Delta) = \mathbb{1}[\text{apply}(\Delta, B_{s\to t})] \wedge \bigwedge_{v \in V_{hidden}} \text{pass}(v, p_\Delta). \tag{4}$$

The indicator is zero when patch application fails. Successful candidates applied only to permitted files and passed all hidden behavioral checks. Unchanged responses, parse errors, timeouts, and missing patches remained in their assigned denominators; no symbolic score substituted for $Y$. An unchanged revision retained its parent's code and was evaluated normally. These targeted assertions neither establish universal correctness nor exclude collateral breakage.

Figure 1 summarizes the isolation and evaluation flow.

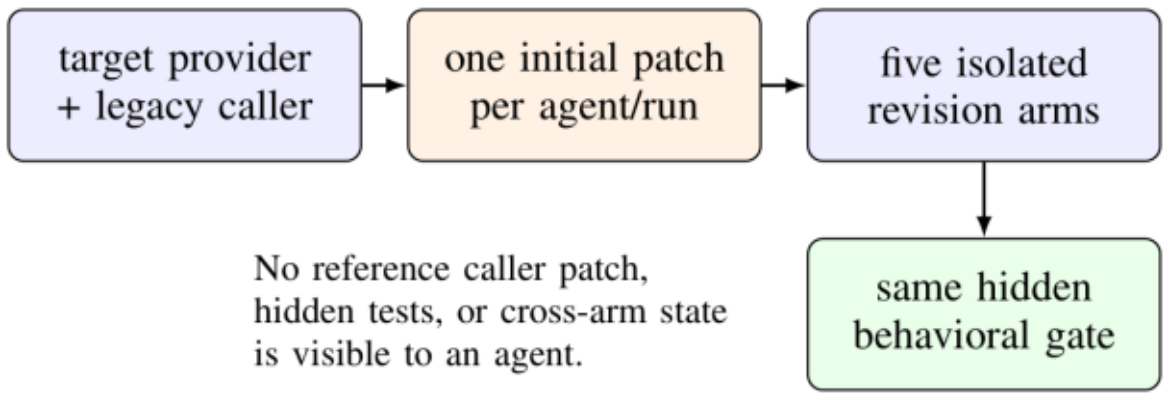

Fig. 1. Evaluation protocol. All revision arms share their exact initial candidate and are evaluated by an independent hidden behavioral outcome.

## III. Controlled Study

### *A. Task Admission and Evidence Separation*

The corpus contained 48 internal migrations, six per repository across eight Python projects: Flask, Pytest, Pluggy, Django, Pydantic, Scrapy, Sphinx, and SQLAlchemy. It included 10 renames, 10 signature changes, 10 relocations, nine protocol changes, and nine deprecations. Task admission followed an eligibility screen that separates provider evidence from sealed caller patches and hidden tests.

Eligible tasks changed non-generated in-repository caller code against an evolving internal interface. Pure documentation edits, test-name changes, duplicate caller migrations from the same provider event, and cases already correct without a client change were excluded. Events and derived templates from the formative pilot were excluded. The task set was frozen after admission; no task was replaced in response to model-run outcomes.

Separate roles enforced isolation. The curator reconstructed $B$ by reverting only client adaptation hunks and sealed the reference target patch. The contract auditor saw $B$, the request, and the provider evidence, but neither the reference patch nor hidden tests. The hidden-test author used the request and target provider interface, without model outputs or feedback contracts. An admission checker then confirmed that the reference patch passed, the legacy and no-op candidates failed for a migration-specific reason, unrelated smoke tests passed, and at least one incomplete migration was rejected. Test failures caused only by a broken environment did not qualify as migration tasks. The independent tests are behavioral assertions, not searches for the same required strings.

All agents received a clean target-provider/legacy-caller snapshot. The complete target client checkout was never substituted for $B$. Git history, sealed patches, network access, hidden-test files, and other arms' outputs are inaccessible. Only the declared caller files were editable. Containers, source and target commits, evidence packets, prompts, and patch-application commands were fixed before model execution.

### B. Agents, Replication, and Budget

The evaluation used Claude Code with Claude Opus 4.6 and Codex with `gpt-5.6-sol`. Each task had two independently generated initial candidates per agent. Client and container configurations were held fixed within each backend across revision arms, and no generation seed was configured. Run order was randomized across conditions and repositories.

Initial generation received the maintenance request, legacy caller, and a fixed target-provider context assembled before runs, without the source diff or temporal checklist. This target context was identical across agents and repetitions and came only from the permitted provider files. The revision experiment added the provider evolution packet according to the arm below; there is no separate retrieval-quality claim or mixed-context treatment.

The accounting is exact: 48×2×2=192 initial candidates, followed by five independent revisions of each, for 960 revisions and 1,152 agent rollout attempts. A rollout could contain several internal model and tool calls, so rollout counts are distinct from API-call and token totals. All cases were revised, including initially successful candidates. Every arm started from a fresh snapshot with the same initial patch applied, so passes, regressions, and non-actionable outputs can be compared under the same policy. Hidden outcomes were never used to select repair cases. For an initially inapplicable patch, all five arms retained $B$ and received the same raw patch and application error. The harness exported each final candidate as a net patch relative to $B$ and evaluated that patch on a fresh copy of $B$; it did not apply an initial change twice.

Each rollout had a ten-minute timeout, a cumulative 16,000 input-token ceiling over all internal model turns (including repeated context and tool results), and a cumulative 6,000 generated-token ceiling. The latter includes reasoning tokens where exposed by the backend accounting interface. All admitted packets fit the common allowance and were not truncated differently by arm. The comparison permitted one revision rollout only.

All arms used the same mediated read/edit tools over permitted files, with no shell, interpreter, test execution, or unrestricted filesystem access. Test files were not mounted in the agent sandbox. An external harness ran the development checks on the initial candidate and supplied their frozen output only to the four test-feedback arms. Retry therefore could not independently obtain new test diagnostics. This evaluates externally supplied feedback under constrained tool access, not unrestricted end-to-end agents. Every attempted rollout remained in its assigned denominator; model, application, timeout, and unresolved infrastructure failures counted as unsuccessful outcomes. Contract authoring and independent audit were additional human inputs beyond benchmark construction. We did not measure their labor cost, so the success-rate comparison does not establish cost effectiveness.

### C. Five Paired Revision Arms

**Retry** received the request and previous patch and was asked to correct it, without new diagnostics. **Tests** additionally received failures from the same frozen development checks. **Diff+tests** received those failures and the complete provider-only version evidence packet and was instructed to derive the migration obligations from it. It was the main direct-evidence control, not a deliberately version-unaware baseline.

**Required+tests** received the same packet and development failures plus the audited missing-required diagnostic $M(p)$. **Full+tests** additionally received $L(p)$, the forbidden-legacy diagnostic. The latter two shared the same audited required set, item ordering, prompts, and thresholds. Removing forbidden feedback does not change task eligibility or other evidence. The full method had no source of knowledge beyond the packet given to the direct-evidence control. Each contract item is traceable to an included anchor; items requiring inaccessible target-client knowledge are excluded by the evidence-separation rule in Section III-A.

The same initial development failures were used for every arm of a parent. Arm order was randomized, and execution occurred in isolated copies. Environment failures received one deterministic infrastructure retry and remained in the denominator if unresolved; model failures were not retried. Aggregation included all assigned outcomes.

### D. Analysis and Diagnostic Challenges

The analysis unit is the migration task. The two agents and two initial runs stay together within that task; 192 parent candidates are not 192 independent migrations. Two prespecified contrasts are full minus diff+tests and full minus required+tests. We average the four paired binary differences per task and then average across tasks. The estimand is this equally weighted, fixed corpus of selected migrations, not an unspecified population of repositories. We use 20,000 hierarchical bootstrap samples, resampling repositories and then tasks within each selected repository while preserving all paired agent/run outcomes. These intervals describe stability under the empirical resampling model. For the two prespecified contrasts, 97.5% per-contrast intervals apply a Bonferroni adjustment for 95% family-wise coverage; ordinary 95% intervals are descriptive. Neither the family-wise level nor 20,000 resamples establishes population coverage with eight purposively selected repositories.

A prespecified sensitivity analysis treats the eight repository mean effects as equally weighted observations and uses a Student $t$ interval with seven degrees of freedom at the same 97.5% per-contrast level. This interval depends on a small-sample $t$ model and does not guarantee nominal coverage with eight repositories. We classify a component as exploratory when its direction is sensitive to the two uncertainty analyses. An inconclusive forbidden contrast does not invalidate a robust package-level main effect.

We report effects for each agent and initial repetition, all repository effects, regressions, and delete-one-repository sensitivity. These analyses do not train on a fold or establish unseen-repository generalization. The paired win/loss counts include unfavorable outcomes and regressions. The post-freeze challenge set further examines valid alternatives flagged by the matcher.

After freezing the contracts and tests, we constructed a targeted challenge set containing one behaviorally invalid but string-compliant candidate and one behaviorally valid alterna-tive implementation per task, for 96 checks. The invalid variants retained all required strings without performing the intended behavior, for example through inert mentions or incomplete state updates. The valid variants exercised alternative spellings or aliases. We evaluated these purpose-built challenges after the freeze and did not use them to tune the contracts or tests. The previously frozen behavioral gate supplied the reported outcomes. This targeted falsification check does not measure the prevalence of either pattern in ordinary agent outputs.

## IV. Results and Interpretation

All revision arms share the same 192 initial candidates. We report hidden-test success, paired contrasts, repository sensitivity, regressions, and diagnostic agreement. Confidence intervals follow the task- and repository-level analysis specified above.

### A. Independent Execution and the Strong Control

The hidden-test successes are 72/192 (37.5%) initially, 100/192 (52.1%) after retry, 112/192 (58.3%) after test feedback, 120/192 (62.5%) after diff+tests, 130/192 (67.7%) after required+tests, and 148/192 (77.1%) after full+tests (Figure 2). All arms have identical denominators and parent candidates. Success is measured by the independent behavioral gate rather than symbolic compliance.

The main contrast is +14.6 percentage points over diff+tests. In the paired outcomes, full wins on 34 parents and loses on six; the remaining outcomes agree. The hierarchical bootstrap 95% interval is [6.3,24.0] points, and the Bonferroni-adjusted 97.5% per-contrast interval is [5.2,25.0]. The repository-*t* sensitivity interval is [6.5,22.6] points at the same per-contrast level. This supports an advantage of the audited diagnostic package over giving an agent the same raw provider history and test failures. The 39.6-point initial-to-full change alone would not identify that advantage.

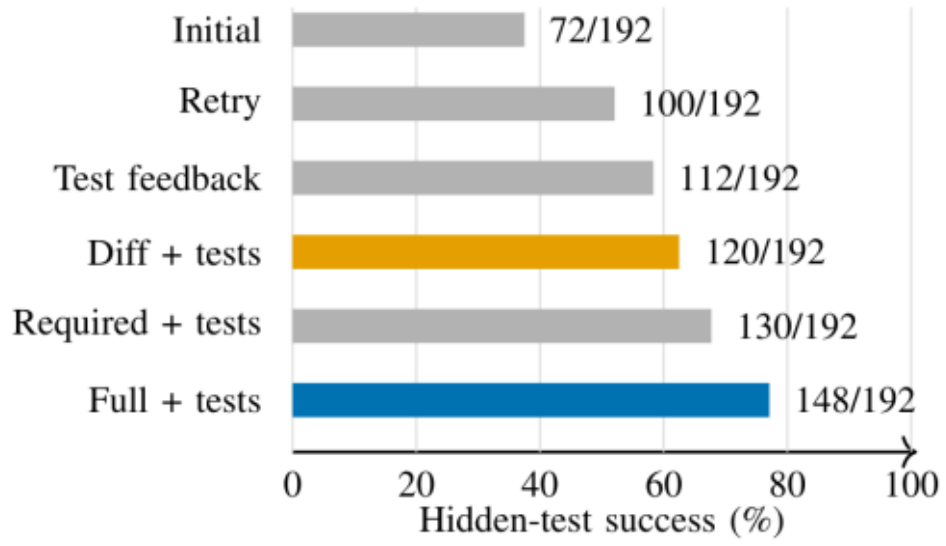

Fig. 2. Hidden-test success for the initial candidates and five paired revision arms. Each condition contains 192 candidates; all revision arms branch from the same initial candidates.

Agent-specific full-minus-diff effects ranged from +11.5 to +17.7 points; repetition-specific effects ranged from +13.5 to +15.6 points. All eight repository effects were positive and ranged from +4.2 to +25.0 points. Delete-one-repository effects ranged from +13.1 to +16.1 points. The effect therefore remained positive when any single repository was removed, while transfer beyond this sample remains untested.

### B. Incremental Value and Regression Risk

Removing forbidden feedback reduces the full-method result by 9.4 points. The pairing has 23 full wins and five losses, with a hierarchical bootstrap 95% interval [2.1,18.2] and nominal 97.5% per-contrast interval [1.0,19.8]. Both agent-specific effects were positive, ranging from +8.3 to +10.4 points. Seven of eight repository effects were positive, with a range of -4.2 to +29.2 points; all delete-one-repository aggregate effects remained positive. However, the repository-*t* 97.5% interval is [-0.8,19.6] points, crossing zero. The forbidden increment is therefore *analysis-sensitive and exploratory*: the aggregate gain does not establish robust confirmation of this component across the two uncertainty analyses.

Among the 72 initially successful parents, the counts of regressions are seven for retry, four for tests, four for diff+tests, two for required+tests, and zero for full. All initially successful candidates were revised and included in these counts. Zero observed regressions does not establish the absence of regression risk.

### C. Symbolic Quality and Targeted Stress Tests

A diagnostic cross-tab on the 192 initial candidates contains 63 symbolic-and-executable passes, 11 symbolic passes that fail execution, nine symbolic failures that pass execution, and 109 failures of both. Thus the precision of symbolic satisfaction for execution is 63/74=85.1%. Symbolic satisfaction is therefore an imperfect indicator of behavioral correctness. This diagnostic agreement does not validate automatic contract synthesis.

For the challenge set, all 48 deliberately string-compliant invalid candidates pass the symbolic matcher but fail hidden execution. Among 48 constructed valid alternatives, 16 pass the symbolic matcher and all 48 pass hidden execution. The lexical matcher rejects 32 valid alternatives, whereas the behavioral gate accepts them and rejects all 48 invalid challenges. This purpose-built stress test illustrates the intended distinction; it does not establish test completeness or characterize the prevalence of deceptive patches in ordinary model outputs.

### D. Interpretation

The 14.6-point gain over diff+tests answers RQ1: the audited diagnostic package improves hidden-test success beyond the same provider evidence and development failures. Its direction is consistent across both agents and the repository sensitivity analyses. The comparison measures the combined effect of human interpretation, obligation wording, and candidate-specific checking.

For RQ2, the 9.4-point gain from forbidden feedback is positive in aggregate and for both agents, but its repository-*t* interval crosses zero. The evidence for this component is consequently exploratory. The diagnostic cross-tab and challenge results also show why symbolic compliance cannot replace independent behavioral evaluation: a patch can contain the expected strings and remain incorrect, or behave correctly through an alternative implementation.

## V. Related Work and Contribution Boundary

SWE-bench studies repository issue resolution [7]; RepoCoder and GraphCoder improve context selection through retrieval and code structure [1], [2]. Documentation grounding and dependency constraints address API hallucination [8], [9]. Hallucination Inspector checks Android migration scaffolding against an API knowledge base [6]. We evaluate patched internal Python callers using independently authored behavioral tests; audited provider obligations are an explicit input, without a claim of automatic knowledge-base construction.

API-evolution research studies refactoring and breaking changes [10], [11]. PyMigBench evaluates cross-library Python migrations using developer changes and client tests [12]. RustEvo$^2$ and When LLMs Lag Behind examine evolving APIs and version conflicts [3], [4]; KCoEvo uses cross-version graphs to guide generation [5]. Our contribution is the controlled evaluation of audited post-generation diagnostics at a fixed target revision, with identical initial candidates and a same-packet diff+tests control. This comparison measures the aggregate diagnostic intervention, without isolating graph or retrieval benefits or establishing superiority over KCoEvo or Hallucination Inspector.

Related validation settings include FaultLens, which learns compact behavioral test suites for generated operational programs [13], and RAGScope, which evaluates evidence gates for hallucination triage in retrieval-augmented generation [14]. Here, audited usage diagnostics guide migration repair, while fixed independent hidden tests determine the behavioral outcome.

## VI. Limitations and Conclusion

The study is limited to Python, eight selected repositories, two backends, and two initial repetitions per agent and task. The agents operate through mediated tools with externally supplied test feedback; the results do not establish performance under unrestricted end-to-end tool use. Human auditing is part of the intervention, and its cost limits conclusions about practical productivity. The experiment does not isolate checking from providing an unchecked audited checklist.

Static usage checks can reject equivalent implementations, as Section IV-C demonstrates. Independent behavioral checks reduce that coupling but remain finite. The manually constructed challenge set tests specific failure modes and does not characterize their frequency in ordinary agent outputs. Test-author mistakes, residual benchmark leakage, and environment differences remain threats despite task admission and evidence separation.

Forty-eight tasks and eight repository clusters do not justify universal generalization. Cluster intervals can be unstable with few groups, and both the bootstrap and repository-$t$ analyses depend on their resampling and small-sample models. The positive package-level effect survives these sensitivity analyses; the forbidden increment does not. Backend versions, task selection, and the human-audited contracts further bound reproducibility and transfer.

GhostAPI-V improves hidden-test success from 62.5% with raw version evidence and development-test feedback to 77.1% with full audited diagnostics on this corpus. This supports a bounded empirical claim for the complete diagnostic package. The additional forbidden-usage component yields a 9.4-point gain over required-only feedback, with analysis-sensitive uncertainty. Independent behavioral evaluation remains essential because symbolic satisfaction alone does not establish a correct migration.